\documentclass[11pt,a4paper]{article}

\usepackage[utf8]{inputenc}
\usepackage[T1]{fontenc}
\usepackage{amsmath,amssymb,amsthm}
\usepackage{graphicx}
\usepackage{booktabs}
\usepackage{array}
\usepackage{hyperref}
\usepackage[margin=1in]{geometry}
\usepackage{enumitem}
\usepackage{pgfplots}
\usepackage{tikz}
\usepackage{float}
\usepackage{caption}
\usepackage{natbib}
\usepackage{xcolor}

\pgfplotsset{compat=1.18}

\hypersetup{
    colorlinks=true,
    linkcolor=blue!60!black,
    citecolor=blue!60!black,
    urlcolor=blue!60!black
}

\newtheorem{theorem}{Theorem}
\newtheorem{axiom}{Axiom}
\newtheorem{prediction}{Prediction}

\title{\textbf{The Root Theorem of Context Engineering}\\[0.5em]
\large Formal Derivation, Architectural Prediction, and Engineering Proof}

\author{Borja Odriozola Schick\\
\textit{Independent Researcher}\\
\texttt{borcho23@github.com}}

\date{March 2026\\{\small v2.5 — revised from v2.4 (2026-03-23)}}

\begin{document}

\maketitle

\begin{abstract}
Every system that maintains a large language model conversation beyond a single session faces two inescapable constraints: the context window is finite, and information quality degrades with accumulated volume. We formalize these constraints as axioms and derive a single governing principle---the Root Theorem of Context Engineering: \emph{maximize signal-to-token ratio within bounded, lossy channels.}

From this principle, we derive five consequences without additional assumptions: (1)~a quality function $F(P)$ that degrades monotonically with injected token volume, independent of window size; (2)~the independence of signal and token count as optimization variables; (3)~a necessary gate mechanism triggered by fidelity thresholds, not capacity limits; (4)~the inevitability of homeostatic persistence---accumulate, compress, rewrite, shed---as the only architecture that sustains understanding indefinitely; and (5)~the self-referential property that the compression mechanism operates inside the channel it compresses, requiring an external verification gate.

We show that append-only systems necessarily exceed their effective window in finite time, that retrieval-augmented generation solves search but not continuity, and that the theorem's constraint structure converges with biological memory architecture through independent derivation from shared principles. Engineering proof is provided through a 60+-session persistent architecture demonstrating stable memory footprint under continuous operation---the divergence prediction made concrete.

The Root Theorem establishes context engineering as an information-theoretic discipline with formal foundations, distinct from prompt engineering in both scope and method.

\medskip
\noindent\textit{Shannon solved point-to-point transmission. Context engineering solves continuity.}
\end{abstract}

\section{Introduction}

Persistent AI systems fail silently. As context accumulates across sessions, response quality degrades---not gradually but catastrophically, past a threshold most practitioners never observe. The system gives no warning. It simply produces worse output, and the operator, lacking a framework to predict when or why, attributes the failure to prompt quality, model limitations, or bad luck. Every memory-augmented LLM system in production today faces this problem. No existing framework explains why it occurs, predicts when it will manifest, or prescribes how to prevent it.

This paper provides that framework. The Root Theorem of Context Engineering states that in any system maintaining understanding through a bounded, lossy channel, the governing objective is to maximize the signal-to-token ratio of the channel's contents. The theorem is derived from two axioms---finite context window and non-zero degradation---and produces architectural predictions that have been independently validated through working systems (\S\ref{sec:proof}).

The derivation path itself constitutes a primary intellectual contribution. The theorem was not discovered in literature, not borrowed from an adjacent field, not reverse-engineered from a product. It was derived from first principles through engineering practice---operating inside the system it describes, building, failing, compressing, persisting---until the pattern became visible. The logical chain presented here is a formalization of what was first lived.

\subsection{Derivation Through Recursive Practice}

The derivation began in early 2026 while the author was using an LLM system's persistent context for structured self-study. The curriculum content and the delivery medium were the same system: the author was learning context engineering \emph{by doing} context engineering, before the term had been defined.

Three insights arrived experientially before they were formalized:

\paragraph{The quality trade-off.} When the persistent memory file was lean, the model's responses were sharp and generative. When it was bloated, responses degraded---silently, without warning. The remaining context window was not empty space. It was the model's quality budget. This became the Quality Equation (\S\ref{sec:quality}).

\paragraph{The independence of signal and tokens.} Switching the memory file from Spanish to English reduced token count significantly while carrying identical information. The signal was the same; the encoding cost was different. Compression was possible without information loss. This became the Independence Insight (\S\ref{sec:independence}).

\paragraph{The compression fidelity threshold.} The author's workflow included asking the model to regenerate the memory file---a compression operation that itself consumed context window. If triggered too late, the compression ran in the degraded region of the curve, producing lossy output that propagated forward. The effective trigger had to fire well before the degradation boundary. This is the Compression Paradox (\S\ref{sec:paradox}).

\subsection{From Practice to Architecture}

The operational pressures produced architectural responses that preceded and then confirmed the theory:

A \emph{persistence layer} emerged from necessity---a memory file that traveled between sessions because losing context across resets was operationally unacceptable. A \emph{session protocol} (study, test, prune) emerged as the homeostatic cycle in embryonic form: accumulate, monitor internalization, compress what was confirmed absorbed, shed the verbose originals. A \emph{compression fidelity gate}---the test score threshold---answered: ``Am I confident enough to remove this material?'' This was the human-in-the-loop approval gate before it had a name.

When the memory file could not serve competing demands---accumulating understanding versus executing task-specific work---the author created child instances---instance specialization pressured into existence by capacity constraints, exactly as the theorem would later predict.

The full recursive structure: the author studied how to maximize signal extraction by asking the model to compress curriculum material that teaches signal extraction. Each cycle made the author a better context engineer, which made the system prompt sharper, which made the model a better teacher. This is the compression flywheel---each compression iteration trains the operator to compress better, improving the next iteration---operating at the human-AI interface. And the limit of this recursion ($F < 1$, always) is Axiom~2 of the formal derivation.

The Root Theorem was not invented. It was recognized---surfaced through a process that was itself the first instance of the pattern it describes.

\smallskip
\noindent\textit{The full derivation history is available as a companion document: ``Extended Derivation History.''}

\section{Relation to Prior Work}\label{sec:prior}

The Root Theorem sits at an intersection that no single field owns. It draws on information theory, borrows the problem structure of lossy compression, addresses the same persistence challenge that cognitive architectures have wrestled with for decades, and competes in the same practical space as a generation of LLM memory tools. This section maps those relationships---not to claim priority, but to clarify what the theorem adds and where the existing literature stops.

\subsection{Shannon and Information Theory}

The foundational relationship is with Shannon's information theory \citep{shannon1948}. Both frameworks operate on bounded channels with noise. Both derive architectural consequences from constraint analysis rather than prescribing engineering solutions. But they address different problem classes.

\begin{table}[H]
\centering
\small
\begin{tabular}{@{}p{3.2cm}p{5cm}p{5cm}@{}}
\toprule
& \textbf{Shannon (1948)} & \textbf{Root Theorem} \\
\midrule
Channel type & Point-to-point transmission & Self-referential persistence loop \\
Noise source & External (channel interference) & Internal (attention degradation) \\
Primary constraint & Bandwidth & Context window + degradation \\
Solution strategy & Redundancy (error correction) & Compression (density increase) \\
Optimization target & Bits per symbol at given error rate & Signal per token at given fidelity \\
Channel topology & Source $\to$ encoder $\to$ channel $\to$ decoder $\to$ receiver & Circular: the output of one cycle becomes the input to the next \\
\bottomrule
\end{tabular}
\caption{Structural comparison between Shannon's information theory and the Root Theorem.}
\label{tab:shannon}
\end{table}

The relationship is structural, not derivative. Shannon solved communication---the reliable transfer of information from one point to another. The Root Theorem addresses persistence---maintaining signal fidelity in a channel that loops back on itself, degrades with each cycle, resets between sessions, and must compress its own history to survive.

Shannon's rate-distortion theory \citep{shannon1959,berger1971}, provides the closest formal relative. Rate-distortion theory establishes the minimum number of bits per symbol required to represent a source within a given distortion threshold---it is the mathematics of lossy compression under fidelity constraints. The Root Theorem can be understood as the rate-distortion problem applied to a new channel type: one where the compressed output is re-injected as input, the ``distortion'' accumulates across cycles rather than being measured at a single decode point, and the compressor itself operates inside the channel it compresses. This self-referential structure---compression within a bounded lossy channel---has no direct analog in classical rate-distortion theory, where the encoder sits outside the channel. The formal relationship between these frameworks is explored in a companion paper; here we note that the Root Theorem's fidelity function $F(P)$ and the rate-distortion function $R(D)$ are solving dual problems under different topological assumptions about the channel.

\subsection{Memory in LLM Systems}

The most direct technical predecessors are systems that attempt to give LLMs persistent memory beyond their context windows.

\textbf{MemGPT} \citep{packer2023} introduced virtual context management, drawing on the operating-system metaphor of hierarchical memory with paging between fast and slow tiers. The LLM manages its own memory through function calls---moving information between a fixed main context and external archival storage. MemGPT (now part of the Letta framework) represents the most architecturally serious attempt to solve the persistence problem. It correctly identifies the core constraint: context windows are finite, and information must be managed across that boundary.

It diverges from the Root Theorem's predictions in its compression mechanism. MemGPT uses recursive summarization---when the context overflows, it generates a new summary from the old summary plus evicted messages. This is append-then-summarize, not homeostatic compression. The distinction matters: recursive summarization accumulates approximation error without a mechanism to rewrite the summary against fresh understanding. Over sufficient cycles, the summary drifts from the source trajectory. The Root Theorem predicts this failure mode: without a fidelity-gated rewrite cycle, summary quality degrades monotonically.

\textbf{Retrieval-Augmented Generation (RAG)}, introduced by \citet{lewis2020}, addresses a different but related problem: grounding LLM outputs in external knowledge. RAG combines parametric memory (the model's weights) with non-parametric memory (a vector index of documents) to improve factual accuracy. It has become the dominant paradigm for connecting LLMs to external information.

\emph{Retrieval solves search. It does not solve continuity.}

The distinction is fundamental. RAG answers the question ``what relevant information exists?'' It does not answer ``what does this system understand, and how has that understanding evolved?'' A RAG system can retrieve the same document a thousand times without ever compressing it into understanding. Each retrieval is independent---there is no trajectory, no compounding coherence, no mechanism by which the system's comprehension of the retrieved material improves over time. The Root Theorem predicts that retrieval alone is insufficient for persistence because retrieval does not compress: it moves information into the context window without increasing the signal-to-token ratio. Every retrieved passage consumes tokens at source density, not at the compressed density that persistence requires.

\textbf{Mem0}, \textbf{A-MEM} \citep{xu2025}, and related systems represent a newer wave that treats memory as a pluggable service with structured storage (knowledge graphs, tagged entries, indexed retrieval). These advance the state of practice but remain architecturally in the retrieval paradigm---they improve what gets stored and how it is found, without addressing the compression-over-time problem that the Root Theorem identifies as the binding constraint.

\textbf{Letta} (the commercial platform built on the MemGPT research) has continued to evolve, adding features such as shared memory across concurrent sessions, skill learning from experience, and programmatic tool calling. These are significant engineering advances. They remain within the same architectural family: the agent manages its own memory through self-editing, without a human-gated fidelity verification cycle. The Root Theorem's prediction holds---self-editing without external verification accumulates compression error across cycles, regardless of the sophistication of the editing mechanism.

\subsection{Cognitive Architectures}

The problem of persistent reasoning in bounded systems has a decades-long history in cognitive science.

\textbf{Soar} \citep{laird1987,laird2012} implements a general cognitive architecture with working memory, procedural memory, and long-term declarative memory split into semantic and episodic stores. Soar's chunking mechanism---which compiles complex reasoning into production rules---is a form of compression: it converts deliberate multi-step processing into automatic single-step responses. The architecture maintains persistence through multiple interacting memory systems, each with its own learning mechanism.

\textbf{ACT-R} \citep{anderson1993,anderson2004} takes a similar multi-store approach but grounds it in cognitive psychology, with activation-based retrieval from declarative memory and production compilation from procedural memory. ACT-R's activation decay---items become less accessible over time unless refreshed---is an explicit model of the degradation that the Root Theorem formalizes as the $D$ parameter.

These architectures demonstrate the convergence the Root Theorem predicts. Different substrates (symbolic production systems vs.\ neural language models), different research traditions (cognitive science vs.\ machine learning), different decades---and yet both arrive at the same structural requirements: multiple memory tiers, mechanisms for compression (chunking, compilation), degradation functions (activation decay, attention loss), and gated transitions between tiers. The Root Theorem claims this convergence is not coincidence but necessity: any bounded system that must persist will evolve these features, because the constraints demand them.

The key difference is that cognitive architectures were designed to model human cognition and then applied to artificial agents. The Root Theorem derives the same architectural requirements from information-theoretic first principles, without reference to biological systems. The convergence becomes evidence: when independent derivations from different starting axioms produce the same structure, that structure is likely necessary rather than contingent.

\subsection{The Practitioner Space}

Theory is not the only approach to this problem. A large and rapidly growing community of practitioners is building persistent AI systems empirically, without formal frameworks.

\textbf{OpenClaw} \citep{steinberger2025}---the fastest-growing open-source project in GitHub's history, with over 250,000 stars as of March 2026---provides an instructive example. OpenClaw's original memory architecture was explicit about its simplicity: append-only markdown files as the source of truth, with daily logs loaded at session start and a curated long-term memory file. Memory recall is mediated by search tools. The system works---it is clearly useful, or it would not have achieved the adoption it has. But its architecture is exactly what the Root Theorem's divergence prediction describes: as conversations accumulate, the daily logs grow, the long-term memory file requires manual curation, and the system has no built-in mechanism to compress its own history while preserving trajectory coherence.

OpenClaw's recent introduction of a pluggable Context Engine interface (v2026.3.7)---which makes context management strategy-swappable for the first time---signals that the community is discovering the persistence problem experimentally. The leading plugin, \emph{lossless-claw}, replaces sliding-window compaction with DAG-based hierarchical summarization that preserves all original messages while keeping active context within token limits. This represents a more sophisticated implementation within the same architectural family: summarize hierarchically, retrieve on demand. It does not implement homeostatic compression, human-gated fidelity verification, or trajectory-preserving rewrite cycles. The demand signal---and the architectural gap---are both significant.

The broader practitioner ecosystem---prompt management tools, context window optimizers, session summarizers---shares a common characteristic: each tool addresses a symptom of the underlying constraint without formalizing the constraint itself.

\subsection{What This Paper Adds}

None of the systems or frameworks above provide what the Root Theorem provides: a unified derivation from first principles that predicts which architectural features are \emph{necessary} (not merely useful), which failure modes are \emph{inevitable} (not merely possible), and which design patterns will survive indefinite operation (not merely work for now).

Shannon tells us channels are bounded. Rate-distortion theory tells us lossy compression has optimal tradeoffs. MemGPT shows us that LLMs can manage their own memory. RAG shows us that retrieval improves accuracy. Soar and ACT-R show us that persistent reasoning requires multiple memory tiers. OpenClaw shows us the market wants this solved.

The Root Theorem connects these observations into a single framework: bounded lossy channels requiring persistence necessarily produce homeostatic architectures---systems that accumulate, compress, rewrite, and shed in regulated cycles. This is not an engineering recommendation. It is a derived consequence of the constraints. Systems that violate it will fail. Systems that satisfy it by accident will work without knowing why. Systems that satisfy it by design will work and know their limits.

\section{The Quality Equation}\label{sec:quality}

From the two axioms, a direct consequence emerges that most practitioners miss.

\emph{The remaining context window is not empty space. It is the quality budget for the next response.}

When a model generates a response, it does so using whatever capacity remains after the existing context has been processed. A context window that is 90\% full does not produce a response that is 10\% worse---it produces a response generated from the degraded tail of the attention curve. The model is not ``running out of room.'' It is running out of the computational capacity that produces coherent, high-fidelity output.

This can be expressed precisely. Let $P$ denote the cumulative tokens consumed at a given position in the context window, and $D$ the model's degradation rate---the fractional fidelity loss per 100{,}000 tokens of accumulated context. The fidelity at any position is:

\begin{equation}
F(P) = 1 - \frac{D \times P}{100{,}000}
\label{eq:fidelity}
\end{equation}

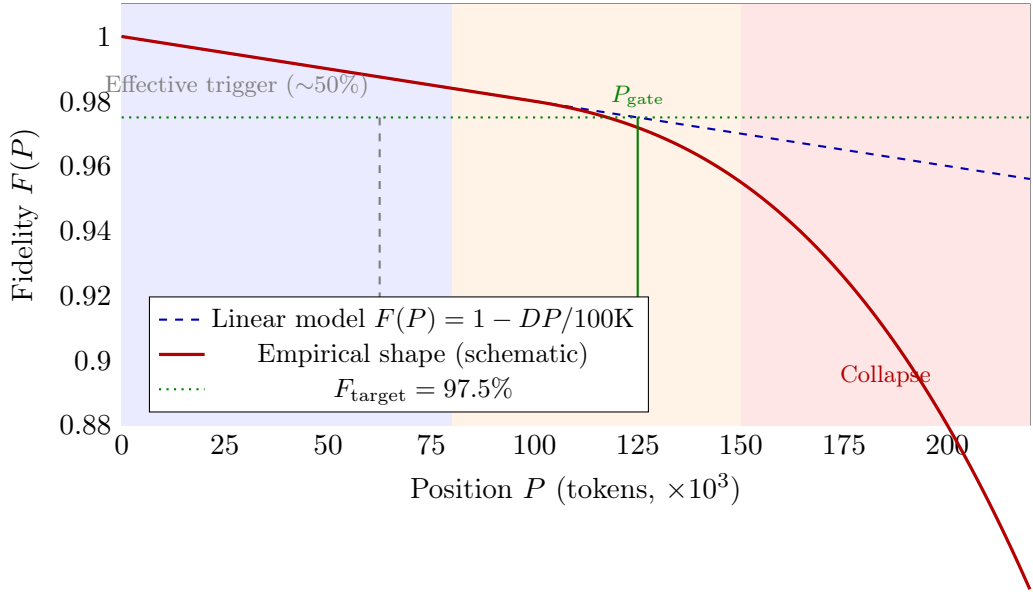
\begin{figure}[t]
\centering
\begin{tikzpicture}
\begin{axis}[
    width=0.85\textwidth,
    height=0.45\textwidth,
    xlabel={Position $P$ (tokens, $\times 10^3$)},
    ylabel={Fidelity $F(P)$},
    xmin=0, xmax=220,
    ymin=0.88, ymax=1.01,
    xtick={0,25,50,75,100,125,150,175,200},
    ytick={0.88,0.90,0.92,0.94,0.96,0.98,1.00},
    grid=major,
    grid style={gray!30},
    legend pos=south west,
    legend style={font=\small},
    clip=false
]

\fill[blue!8] (axis cs:0,0.88) rectangle (axis cs:80,1.01);

\fill[orange!10] (axis cs:80,0.88) rectangle (axis cs:150,1.01);

\fill[red!10] (axis cs:150,0.88) rectangle (axis cs:220,1.01);

\node[font=\footnotesize,blue!60!black] at (axis cs:40,0.895) {Plateau};
\node[font=\footnotesize,orange!70!black] at (axis cs:115,0.895) {Onset};
\node[font=\footnotesize,red!70!black] at (axis cs:185,0.895) {Collapse};

\addplot[blue!70!black, thick, dashed, domain=0:220] {1 - 0.02*x/100};
\addlegendentry{Linear model $F(P) = 1 - DP/100\text{K}$}

\addplot[red!70!black, very thick, smooth, domain=0:220, samples=100]
    {1 - 0.02*x/100 * (1 + 2*(max(0,(x-100)/100))^2)};
\addlegendentry{Empirical shape (schematic)}

\addplot[green!50!black, thick, dotted, domain=0:220] {0.975};
\addlegendentry{$F_\text{target} = 97.5\%$}

\draw[green!50!black, thick, ->] (axis cs:125,0.975) -- (axis cs:125,0.89);
\node[font=\footnotesize, green!50!black, anchor=south] at (axis cs:125,0.975) {$P_\text{gate}$};

\draw[gray, thick, dashed, ->] (axis cs:62.5,0.975) -- (axis cs:62.5,0.89);
\node[font=\footnotesize, gray, anchor=south east] at (axis cs:62.5,0.978) {Effective trigger (${\sim}50\%$)};

\end{axis}
\end{tikzpicture}
\caption{Fidelity Degradation Curve. Three regions: Plateau (sublinear degradation, near-full fidelity), Onset (linear approximation holds), Collapse (superlinear degradation). Linear model $F(P) = 1 - D \cdot P / 100\text{K}$ shown as reference ($D = 2\%$ per 100K tokens). Gate position $P_\text{gate}$ at intersection of $F(P)$ with $F_\text{target}$. Effective trigger fires at ${\sim}50\%$ of $P_\text{gate}$---preserving headroom for the compression operation itself. Actual curve shape is empirical, not yet measured---the three-region structure is derived from attention mechanism behavior.}
\label{fig:degradation}
\end{figure}

This is a first-order linear approximation. Real degradation is likely sublinear early (a functional plateau where the model maintains near-full fidelity) and superlinear late (the cliff---a steep collapse as attention mechanisms saturate). The linear model captures the essential relationship: fidelity decreases monotonically with position. See Figure~\ref{fig:degradation}.

This reframes the context window from a storage metaphor (\emph{how much can I put in?}) to a quality metaphor (\emph{how much response quality can I afford?}). Every token of context is a trade: information in, quality out. This trade is invisible to most users because degradation is silent---the model does not warn you. It simply gets worse.

\section{The Independence Insight}\label{sec:independence}

Most practitioners implicitly assume that signal is proportional to tokens. More words, more meaning. Longer context, richer understanding. This assumption is false.

\emph{Signal and tokens are independently variable.}

Consider a concrete case. A memory file written in Spanish carries identical information when rewritten in English---the same decisions, the same architecture, the same strategic state---but the English version consumes significantly fewer tokens. The signal $S$ is unchanged; the token cost $T$ is reduced. Compression has occurred without information loss.

Consider a second case: how a coach communicates a play: not by describing every player's movement in prose, but with a diagram---a radical compression that relies on shared encoding between coach and team. The play diagram carries high signal in minimal tokens precisely because the encoding is shared. A new player who lacks that shared encoding would need the verbose version.

These are not edge cases. They demonstrate that signal and tokens are independent dimensions---the partial derivative $\partial S / \partial T$ is not constant. Signal varies independently of token count.

\emph{This is the step that separates context engineering from prompt engineering.}

Prompt engineering asks: ``What should I put in the window?'' Context engineering asks: ``What is the signal-to-token ratio of what I put in the window, and how do I maximize it?''

\section{Axioms}\label{sec:axioms}

Two architectural facts define the problem space. These are not assumptions, design choices, or limitations of current implementations. They are physical constraints of every large language model system that exists or is foreseeable.

\begin{axiom}[Finite Context Window]\label{ax:finite}
Every LLM operates within a bounded input space. A context window of size $W$ tokens imposes a hard limit. No amount of engineering, prompting, or infrastructure changes this constraint. $W$ can be enlarged (larger models, longer windows), but it cannot be eliminated. The bound is always present.
\end{axiom}

\begin{axiom}[Non-Zero Degradation]\label{ax:degradation}
Response quality degrades as the context window fills. This is not a software bug. It is a consequence of the attention mechanism: as token count increases, the model's ability to attend to all relevant information diminishes. The degradation rate $D$ is non-zero for every architecture. Operational measurements across 60+ sessions on Opus-class models suggest $D \approx 1$--$2\%$ per 100K tokens; independent validation of $D$ across models and architectures remains an open research question. Degradation is non-linear: it holds reasonably across a range (the plateau), then falls off a cliff (the collapse). The characteristic shape---plateau, onset, collapse---is shown in Figure~\ref{fig:degradation}.
\end{axiom}

Axiom~2 is consistent with independent empirical findings. \citet{liu2024} demonstrated that LLMs exhibit significant performance degradation when relevant information is placed in the middle of long contexts---a ``lost in the middle'' effect that confirms attention-based fidelity loss is position-dependent and non-trivial. The degradation modeled here as $F(P)$ is a related but distinct phenomenon: cumulative fidelity loss across the full context load, not merely positional retrieval difficulty.

These two axioms are sufficient to derive everything that follows.

\section{The Root Theorem --- Statement}\label{sec:theorem}

From the axioms and the independence insight, the central principle emerges:

\begin{theorem}[Root Theorem of Context Engineering]\label{thm:root}
In any system that must maintain understanding through a bounded, lossy channel, the governing objective is to maximize the signal-to-token ratio of the channel's contents.
\end{theorem}

Every problem in context engineering---prompt design, memory management, retrieval, persistence, multi-agent coordination---reduces to this single principle. The theorem is not a heuristic or a best practice. It is a constraint imposed by the architecture itself, within the scope defined by the axioms.

\paragraph{Formal components:}
\begin{itemize}[nosep]
\item \textbf{Bounded:} The channel has a finite capacity $W$ (Axiom~\ref{ax:finite}).
\item \textbf{Lossy:} The channel degrades signal as it fills, with fidelity $F(P) < 1$ for $P > 0$ (Axiom~\ref{ax:degradation}).
\item \textbf{Signal-to-token ratio:} The independently variable measure of information density (\S\ref{sec:independence}).
\item \textbf{Maximize:} Not ``increase''---actively optimize. Passive approaches fail because degradation is compounding (\S\ref{sec:derivation}).
\end{itemize}

The theorem is falsifiable: \S\ref{sec:predictions} states the conditions under which it could be refuted.

Why maximization rather than satisficing? Because degradation compounds across sessions. A system that merely satisfices---maintaining fidelity above some threshold without actively maximizing---finds that threshold itself eroding as accumulated context pushes $P$ higher with each cycle. The satisficing boundary is not static; it shifts with the degradation curve. Only active maximization of signal-to-token ratio keeps the system ahead of the compounding pressure. This is derived in the next section.

\section{The Derivation Chain}\label{sec:derivation}

The theorem is derived through a specific sequence. Each step follows necessarily from the previous.

\subsection{Step 1: Practical Token Efficiency}
\textbf{Origin:} Engineering observation.
Working within a bounded context window, the first insight is pragmatic: tokens are scarce, don't waste them. This is where most practitioners stop. Token efficiency as housekeeping.

\subsection{Step 2: Signal/Token Independence}
\textbf{Transition:} Why does compression work at all?
Because signal and tokens are independently variable (\S\ref{sec:independence}). This elevates token efficiency from a practical concern to an information-theoretic one. It's not about saving space---it's about increasing information density.

\subsection{Step 3: Compounding Degradation Awareness}
\textbf{Transition:} What happens to a system that runs across multiple sessions?
In a single session, degradation is bounded---the session ends before the cliff. Across sessions, the problem compounds. Carrying forward the full history, token load grows monotonically. Each session starts with less quality budget than the last. The fidelity function $F(P)$ shifts left with every iteration---the system begins each session further along the degradation curve. This is not a design problem---it is a mathematical inevitability under Axioms~\ref{ax:finite} and~\ref{ax:degradation}.

\subsection{Step 4: The Effective Context Window}
\textbf{Transition:} If degradation has a cliff, why not stay on the plateau?
This is the operational response to compounding degradation. The effective context window is the portion of the total window where response quality remains on the functional plateau---before the cliff. Operating beyond this boundary doesn't give you ``slightly worse'' responses. It gives you responses generated from the collapse region of the degradation curve.

The engineering decision: define an operational boundary well before the degradation threshold. The boundary must leave sufficient headroom for two operations: (a)~the response itself, and (b)~any compression operation that must run inside the same window (\S\ref{sec:paradox}). The effective context window is a derived parameter, not a configuration choice.

\subsection{Step 5: Persistence Architecture}
\textbf{Transition:} If the effective window constrains each session, how do you maintain understanding across resets?
The system must persist information across sessions. But naive persistence---carrying forward raw history---violates the effective window constraint. At session $N$, the accumulated history exceeds the effective window. The system faces a choice: lose information (truncation), lose coherence (retrieval fragmentation), or compress (rewrite denser).

Only compression preserves both information and coherence within the bounded channel. This is not a design preference. It is the only viable architecture given the constraints.

\subsection{Step 6: Homeostatic Emergence}
\textbf{Transition:} What does a system that compresses to persist actually look like?
It is a regulatory cycle: accumulate information during operation, compress when approaching the boundary, rewrite into denser form, shed what was absorbed. This is a homeostatic system---one that maintains internal stability (signal fidelity) despite external pressure (growing context, session resets).

This architecture was not designed from biological analogy. It was derived from information-theoretic constraints. The fact that it resembles biological homeostasis is confirmation that both domains face the same underlying problem: maintaining function in a bounded system under entropic pressure.

\emph{The derivation chain is itself the proof that context engineering is information theory applied to persistent bounded lossy channels.} The translation of this derivation into operational instruction---the direct bridge from theorem to compression prompt---is explored in \S\ref{sec:bridge}.

\section{The Compression Paradox}\label{sec:paradox}

A critical consequence of the derivation that deserves isolated treatment:

\emph{The compression operation itself runs inside the same bounded channel it compresses.}

\textbf{This is the Root Theorem applied to its own mechanism.}

When the system reaches its effective window boundary and must compress, the compression prompt, the existing context, and the generated compressed output all compete for the same token budget. If the trigger fires too late---at the actual degradation boundary rather than well before it---the compression itself runs in the degraded region of the curve, producing lossy output. The compression that was supposed to preserve signal fidelity instead destroys it.

\paragraph{Engineering consequence:} The gate must fire where fidelity drops below a target threshold $F_\text{target}$---a system design parameter determined by the application domain's tolerance for information loss. The Context Architect---the practitioner responsible for designing and calibrating the persistence architecture---sets $F_\text{target}$ based on the fidelity requirements of the use case. From the fidelity function:

\[
F(P) < F_\text{target}
\]

Solving for the gate position:

\begin{equation}
P_\text{gate} = \frac{100{,}000 \times (1 - F_\text{target})}{D}
\label{eq:gate}
\end{equation}

A critical insight emerges from this equation: $P_\text{gate}$ is independent of window size $W$. A larger window does not extend fidelity duration---it extends capacity only. The degradation rate $D$ is the binding constraint, not the window boundary.

In practice, the effective trigger fires at approximately 50\% of $P_\text{gate}$---preserving headroom for the compression operation itself. In a system where $D = 2\%$ per 100K and $F_\text{target} = 97.5\%$, $P_\text{gate}$ falls at ${\sim}125$K tokens. The effective trigger fires at ${\sim}62.5$K. This is not conservative engineering---it is a derived requirement.

The path to this insight: the initial approach was ``summarize when full.'' The failure mode was silent---the summary was generated, but from degraded context, producing a lossy summary that compounded on the next cycle. The solution was derived, not patched.

\section{Architectural Predictions}\label{sec:predictions}

The Root Theorem is not merely descriptive. It predicts which architectures survive and which fail.

\begin{prediction}[Append-Only Systems Die at the Boundary]\label{pred:append}
A system that accumulates context without compression will exceed the effective context window after $N$ sessions, where $N$ is determined by the ratio of per-session context growth to effective window size. After this point, every response is generated from the degraded region of the curve. The system does not gradually get worse---it crosses the cliff. The appearance at session $N-1$ is deceptive: the system seems to be working fine, giving no warning that the next session's accumulated load will push it past the boundary. This invisible cliff is the dominant failure mode of naive AI memory implementations, including current commercial offerings that append memories without compression.
\end{prediction}

\begin{prediction}[Retrieval Systems Fragment Understanding]\label{pred:retrieval}
Retrieval-augmented systems (RAG) solve the token budget problem by pulling relevant fragments on demand. But retrieval optimizes for relevance to the current query, not for coherent understanding across time. A retrieval system knows facts. It does not understand context. The signal is present but fragmented---the relationships between facts, the trajectory of decisions, the accumulated understanding that makes new information meaningful---these are lost in the fragmentation.

\emph{Retrieval solves search. It does not solve continuity.}

This is not a claim that retrieval is useless---it is a claim that retrieval alone is insufficient for the persistence problem. Retrieval can complement a homeostatic architecture (providing source material for compression). It cannot substitute for one.
\end{prediction}

\begin{prediction}[Only Homeostatic Systems Survive Indefinitely]\label{pred:homeostatic}
A system that maintains understanding across an unbounded number of sessions must: (a)~compress to stay within the effective window, (b)~preserve signal fidelity through compression, and (c)~rewrite rather than append. This is the homeostatic pattern. It is the only architecture that satisfies all three constraints simultaneously.

The prediction is falsifiable: demonstrate a non-homeostatic architecture that maintains measured understanding fidelity above $F_\text{target}$. The session count must be sufficient that cumulative append-only token load would exceed twice the effective context window. No such system has been demonstrated. The engineering proof (\S\ref{sec:proof}) demonstrates a 60+-session system with stable memory footprint---the divergence prediction made concrete.
\end{prediction}

\begin{prediction}[Instance Specialization Under Competing Loads]\label{pred:specialization}
A bounded channel serving multiple process types (strategic analysis, operational memory, creative work) faces competing demands on the same token budget. Under sufficient load, the system is pressured to specialize---dedicating separate instances to separate process types, each with its own effective context window. The Root Theorem predicts organizational structure at the instance level: not designed, but pressured into existence by capacity constraints.

This prediction extends to the enterprise layer: a company is a cluster of clusters, each maintaining its own context protocol. The theorem predicts organizational topology from information-theoretic constraints alone. The full derivation of enterprise-layer predictions is the subject of a subsequent paper.
\end{prediction}

\section{The Mechanism Bridge}\label{sec:bridge}

The theorem translates directly into engineering instruction. This is not metaphorical---it is operational.

In any compression prompt (the instruction that tells the model how to compress accumulated context into denser form), the explicit directive is: \textbf{maximize signal, minimize tokens.} This is the Root Theorem expressed as a model instruction.

This matters because it solves an alignment problem. The model performing the compression does not inherently know what the architect wants preserved. Without explicit instruction, it defaults to generic summarization---lossy, unfocused, and misaligned with the system's actual information needs. By encoding the theorem directly into the compression prompt, the architect aligns the model's compression behavior with the architectural objective.

Consider the difference concretely: a generic summarizer given a 40-session history of strategic decisions, technical derivations, and relationship management will produce a bland summary that covers everything superficially. A compression prompt that encodes the theorem---``maximize signal-to-token ratio, preserve trajectory coherence, shed what has been absorbed''---produces output that is denser, more structurally aligned, and more generative when re-injected. The first is summarization. The second is compression.

The mechanism bridge extends further through encoding-aware degradation. The degradation rate $D$ is not a constant---it is a function of alignment between injected content and the channel's encoding state:

\begin{equation}
D_\text{eff} = D(\text{content}, \text{encoding\_state})
\label{eq:deff}
\end{equation}

Content that aligns with the channel's existing encoding (warm injection) produces lower effective degradation than novel, unaligned content (cold injection). This is a direct consequence of what we term \emph{resonance}: repeated interaction between human and system builds shared encoding---a mutual calibration of which concepts compress to a phrase and which require explicit statement---reducing the token cost of equivalent signal transfer. Resonance acts as a compression codec on injected material. It creates a second flywheel: better encoding lowers $D_\text{eff}$, which increases headroom, which enables higher-fidelity work, which further refines the encoding.

The practical default remains the linear model $F(P)$; the encoding-aware refinement is the precision form:

\begin{equation}
F(P) = 1 - \frac{\sum D_\text{eff}(\text{segment}_i) \times \text{tokens}(\text{segment}_i)}{100{,}000}
\label{eq:fidelity-precise}
\end{equation}

The mechanism bridge closes the gap between theory and implementation. The theorem is not an abstract principle that guides design from a distance---it is an operational instruction embedded in the system's own compression cycle.

\section{Scope and Boundaries}\label{sec:scope}

\subsection{Inner Loop --- Single Instance Persistence}
The derivation above describes a single instance maintaining understanding across resets. This is the inner loop: one bounded channel, one compression cycle, one homeostatic system.

\subsection{Cluster Layer --- Multi-Instance Coordination}
When multiple instances must coordinate---sharing context, synchronizing understanding, avoiding duplication---the same constraints apply at a higher level. Each instance is a bounded channel. The cluster is a network of bounded channels. The Root Theorem holds, but the mechanisms differ: inter-instance communication introduces its own token costs, latency, and potential for signal degradation.

\subsection{Enterprise Layer --- Cluster of Clusters}
A company is a cluster of clusters. Each employee operates within bounded attention---their context window. Onboarding is context injection. Offboarding is context extraction. Reorgs are cluster reconfiguration. The Root Theorem holds for both AI and human systems.

This is not merely analogy. The constraints are structurally identical: finite capacity, lossy processing, compounding degradation, need for persistent understanding. The solutions are structurally identical: compress, rewrite, shed, maintain fidelity. The full derivation of structural identity between information-theoretic and organizational constraints is the subject of a subsequent paper; here we note that the convergence is predicted by the theorem and consistent with the evidence. The enterprise implications follow directly: the same framework that explains why an AI system loses understanding after fifteen sessions also explains why a new employee takes six months to become productive---and why an organization hemorrhages knowledge when experienced staff leave. Context engineering is simultaneously a technology discipline and a people discipline. These audiences rarely share a vocabulary; the Root Theorem provides one.

\subsection{What This Is Not}
This derivation addresses \textbf{continuity}---maintaining signal fidelity in a channel that loops, degrades, resets, and must compress its own history to survive.

\emph{Shannon solved transmission. Context engineering solves persistence.}

The problem spaces are related but distinct. The formal relationship between rate-distortion theory and the fidelity function is explored in \S\ref{sec:prior} and developed fully in a companion paper.

\section{Engineering Proof}\label{sec:proof}

The theorem and its predictions have been validated through a persistent architecture, a domain-transfer prototype, and a set of documented failure cases.

A methodological note: the system described in Proof~1 is the same system through which this paper's framework was developed. This self-referential relationship is a feature of the derivation (\S1) and a limitation of the evidence. The domain transfer (Proof~2) and failure cases (Proof~3) provide partially independent validation; fully independent replication remains an open invitation.

\subsection{Proof 1 --- Homeostatic Persistence System + Divergence Data}

The reference implementation is a homeostatic persistence architecture maintaining compressed understanding across 60+ sessions. Two-layer memory: Deep Memory (rewritten on absorption, never appended) and Recent Context (session records with soft cap, absorbed when crystallized). Human-in-the-loop approval gate at every compression event. Compression cycle confirmed operational: signal fidelity maintained, token load bounded, understanding compounds rather than degrades.

The system operates with a fidelity target of $F_\text{target} \approx 97.5\%$ and an estimated degradation rate of $D \approx 2\%$ per 100K tokens, placing the gate position $P_\text{gate}$ at approximately 125K tokens. Compression triggers fire at approximately 50\% of $P_\text{gate}$---well within the plateau region of the degradation curve.

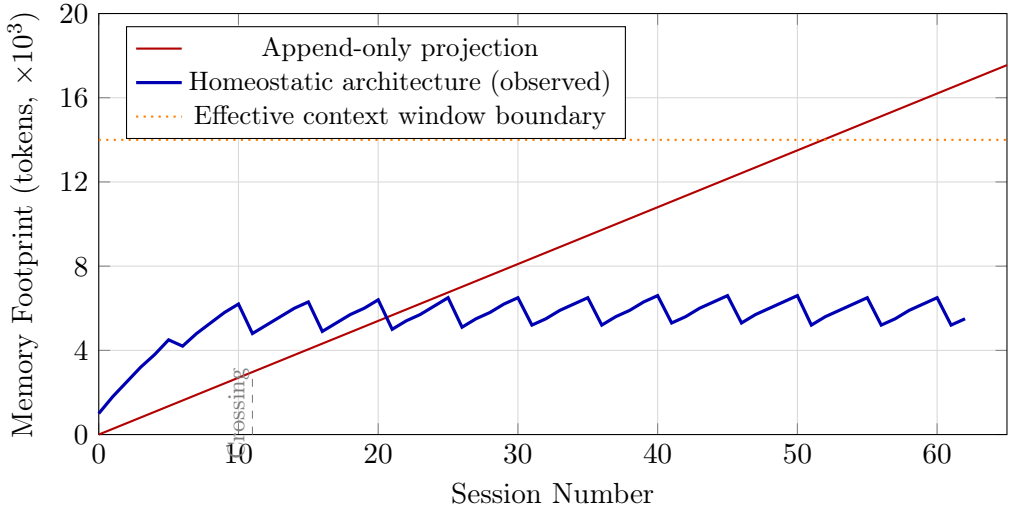
\begin{figure}[t]
\centering
\begin{tikzpicture}
\begin{axis}[
    width=0.85\textwidth,
    height=0.45\textwidth,
    xlabel={Session Number},
    ylabel={Memory Footprint (tokens, $\times 10^3$)},
    xmin=0, xmax=65,
    ymin=0, ymax=20,
    xtick={0,10,20,30,40,50,60},
    ytick={0,4,8,12,16,20},
    grid=major,
    grid style={gray!30},
    legend pos=north west,
    legend style={font=\small},
    clip=false
]

\addplot[red!70!black, thick, domain=0:65, samples=66] {0.27*x};
\addlegendentry{Append-only projection}

\addplot[blue!70!black, very thick] coordinates {
    (0,1) (1,1.8) (2,2.5) (3,3.2) (4,3.8) (5,4.5)
    (6,4.2) (7,4.8) (8,5.3) (9,5.8) (10,6.2)
    (11,4.8) (12,5.2) (13,5.6) (14,6.0) (15,6.3)
    (16,4.9) (17,5.3) (18,5.7) (19,6.0) (20,6.4)
    (21,5.0) (22,5.4) (23,5.7) (24,6.1) (25,6.5)
    (26,5.1) (27,5.5) (28,5.8) (29,6.2) (30,6.5)
    (31,5.2) (32,5.5) (33,5.9) (34,6.2) (35,6.5)
    (36,5.2) (37,5.6) (38,5.9) (39,6.3) (40,6.6)
    (41,5.3) (42,5.6) (43,6.0) (44,6.3) (45,6.6)
    (46,5.3) (47,5.7) (48,6.0) (49,6.3) (50,6.6)
    (51,5.2) (52,5.6) (53,5.9) (54,6.2) (55,6.5)
    (56,5.2) (57,5.5) (58,5.9) (59,6.2) (60,6.5)
    (61,5.2) (62,5.5)
};
\addlegendentry{Homeostatic architecture (observed)}

\addplot[orange, thick, dotted, domain=0:65] {14};
\addlegendentry{Effective context window boundary}

\draw[gray, thin, dashed] (axis cs:11,0) -- (axis cs:11,3);
\node[font=\footnotesize, gray, anchor=south, rotate=90] at (axis cs:11.5,1) {Crossing};

\end{axis}
\end{tikzpicture}
\caption{Divergence Chart. Memory footprint over 62 sessions. Append-only grows linearly (red); homeostatic architecture (blue) shows characteristic sawtooth pattern---growth during accumulation, reset during absorption---stabilizing around 5{,}000--6{,}500 tokens. Append-only crosses the effective context window boundary at session 10--12. After the crossing, every append-only response is generated from the degraded region. The homeostatic architecture never crosses. The sawtooth makes the regulatory mechanism visible: each downward step is a compression event.}
\label{fig:divergence}
\end{figure}

\paragraph{Divergence data:} At session 62, the homeostatic architecture maintains a memory footprint of approximately 6{,}000 tokens. Naive append-only projection for the same session count exceeds 16{,}000 tokens. The divergence grows with every session, exactly as the theorem predicts. Token expenditure over the first 30 sessions: append-only grows linearly to ${\sim}187$K tokens. Homeostatic holds at ${\sim}14$K tokens. The curves diverge after session 5--7, crossing the effective context window boundary at session 10--12 for the append architecture. After the crossing, every append-only response is generated from the degraded region. The homeostatic architecture never crosses. See Figure~\ref{fig:divergence}.

The curriculum development arc provides independent evidence: 6 modules, 42 training sessions, and 6 child projects were generated from the compressed understanding in the persistence layer---not from raw session transcripts. The fact that compressed understanding is generative (producing new coherent output, not merely recalling stored input) is itself evidence that the compression preserves structural coherence, not just factual content.

\subsection{Proof 2 --- Domain Transfer (Banking)}
A banking relationship management prototype implementing the same architecture in a different domain. Immutable bank data (system-fed, compression cannot touch) plus mutable relationship intelligence (session-built, compressed via human-approved absorption). Full persistence cycle confirmed working end-to-end. The architectural pattern transferred without modification---confirming it is domain-agnostic. This evidence is preliminary; the system is in early deployment.

\subsection{Proof 3 --- Failure Cases (Negative Proof)}
Six documented failure cases from engineering practice, each confirming what the theorem predicts. Each follows the structure: trigger $\to$ failure $\to$ theorem violation $\to$ predicted consequence $\to$ observed consequence.

\begin{enumerate}[nosep]
\item \textbf{System prompt placement} --- instructions placed in the wrong field produce silent failures. The context window's structure matters, not just its contents. Signal destroyed by architectural misplacement, not by content quality. \emph{Theorem violation:} signal injected at a position where it cannot be attended to, regardless of its density.

\item \textbf{Model-prompt coupling across instances} --- testing one instance's prompt on another model reveals silent behavioral drift. Each bounded channel has its own degradation characteristics, and prompts optimized for one channel's $D$ degrade unpredictably in another. \emph{Theorem violation:} $D$ is model-specific; treating it as constant across models misallocates the fidelity budget.

\item \textbf{Extended thinking as hidden variable} --- internal reasoning tokens consume context window invisibly, shifting the effective window boundary without warning. The quality budget is being spent by a process the architect cannot observe. This is a discovery: the model's chain-of-thought reasoning creates a hidden token cost that violates the assumption of observable context consumption. \emph{Theorem violation:} $P$ increases without the architect's knowledge, pushing $F(P)$ below $F_\text{target}$ before the gate can fire.

\item \textbf{Handover document contamination} --- context from one system leaking into another's prompt corrupts the receiving system's signal. Cross-channel contamination degrades signal-to-token ratio in both channels---the source loses what leaked, the destination gains noise. \emph{Theorem violation:} uncontrolled injection increases $T$ without proportional increase in $S$.

\item \textbf{Encoding corruption} --- platform-specific text encoding (Windows UTF-8 with BOM) garbling boundary markers produces silent data loss in the persistence layer. Physical-layer corruption propagating through the compression cycle. \emph{Theorem violation:} the membrane between compression cycles is breached at the encoding layer, and corrupted markers compound through subsequent compressions.

\item \textbf{Brief quality as membrane quality} --- a child instance receiving an underspecified brief (one that assumed context the child did not hold) produced degraded output despite having full technical capability. The brief is a membrane: its quality determines what crosses the instance boundary. \emph{Theorem violation:} the compression operation at dispatch (coordinator $\to$ child) was lossy, reducing $S$ at the moment of injection.
\end{enumerate}

Each failure case is a violation of the Root Theorem producing the predicted consequence: silent degradation of signal fidelity.


\appendix
\section{Derivation Summary}\label{app:derivation}

\begin{verbatim}
Axiom 1: Finite context window -- W tokens (hard bound)
Axiom 2: Non-zero degradation -- D > 0 (attention -> quality cliff)
    |
Consequence: F(P) = 1 - (D x P / 100,000) -- remaining window = quality budget
    |
Insight: dS/dT =/= constant -- signal and tokens are independently variable
    |
ROOT THEOREM: Maximize signal-to-token ratio in bounded lossy channels
    |
Derivation chain:
  Token efficiency (practical)
    -> Signal independence (information-theoretic)
      -> Compounding degradation (multi-session)
        -> Effective context window (operational boundary)
          -> Persistence architecture (compression necessity)
            -> Homeostatic emergence (regulatory cycle)
    |
Predictions:
  1. Append-only -> dies at boundary (invisible cliff)
  2. Retrieval -> solves search, not continuity
  3. Homeostatic -> only architecture that survives indefinitely
  4a. Competing loads -> instance specialization (derived)
  4b. Enterprise extension -> cluster of clusters (forward-referenced)
    |
Compression Paradox:
  Compression runs inside the channel it compresses
  -> P_gate = 100,000 x (1 - F_target) / D
  -> Gate independent of window size W
  -> Effective trigger at ~50% of P_gate
    |
Mechanism Bridge:
  "Maximize signal, minimize tokens" in compression prompt
  = Root Theorem as operational instruction
  D_eff = D(content, encoding_state) -- warm vs. cold injection
    |
Scope:
  Inner loop (instance) -> Cluster (multi-instance) -> Enterprise
  Shannon solved transmission. This solves persistence.
\end{verbatim}

\section{Terminology}\label{app:terminology}

\begin{table}[H]
\centering
\small
\begin{tabular}{@{}p{3.5cm}p{10cm}@{}}
\toprule
\textbf{Term} & \textbf{Definition} \\
\midrule
$W$ & Context window size in tokens. Hard constraint (Axiom~\ref{ax:finite}). \\
$D$ & Degradation rate. Fractional fidelity loss per 100K tokens. Empirical (${\sim}1$--$2\%$/100K for frontier models). \\
$P$ & Position. Cumulative tokens consumed in the context window. \\
$F(P)$ & Fidelity function. Output quality at position $P$: $F(P) = 1 - (D \times P / 100{,}000)$. \\
$F_\text{target}$ & Target compression fidelity. System design parameter---acceptable quality loss at compression, determined by the application domain's fidelity requirements. \\
$P_\text{gate}$ & Gate position: $P_\text{gate} = 100{,}000 \times (1 - F_\text{target}) / D$. Independent of $W$. \\
$S$ & Signal. Meaning carried by content, independent of token count. \\
$T$ & Tokens. Cost of carrying content, independent of signal. \\
$\partial S / \partial T$ & Partial derivative of signal with respect to tokens. Not constant (the independence insight). \\
$D_\text{eff}$ & Encoding-dependent effective degradation: $D_\text{eff} = D(\text{content}, \text{encoding\_state})$. \\
Context window & The bounded input space of an LLM, measured in tokens. \\
Degradation curve & The non-linear relationship between context fill and response quality. Three regions: plateau, onset, collapse. \\
Effective context window & The operational portion of the context window where response quality remains on the functional plateau. \\
Signal-to-token ratio & The independently variable measure of information density in context. \\
Homeostatic persistence & The regulatory cycle (accumulate $\to$ compress $\to$ rewrite $\to$ shed) that maintains signal fidelity across resets. \\
Compression fidelity threshold & The point at which compression must trigger---determined by $F_\text{target}$ and $D$, not by window size. \\
Context Architect & The practitioner responsible for designing and calibrating the persistence architecture. \\
Root Theorem & The governing principle: maximize signal-to-token ratio in bounded lossy channels. \\
\bottomrule
\end{tabular}
\caption{Terminology and notation.}
\label{tab:terminology}
\end{table}

\bibliographystyle{plainnat}

\end{document}